\batchmode
\documentclass{article}

\usepackage{latexsym, amssymb, amsfonts}
\usepackage[dvipdfm]{graphicx}
\usepackage[round]{natbib}
\usepackage{fancyvrb}

\title{Towards an Intelligent Database System Founded on the SP Theory of Computing and Cognition}

\author{J Gerard Wolff \\ CognitionResearch.org.uk}

\begin{document}

\maketitle

\begin{abstract}

The SP theory of computing and cognition, described in previous publications, is an attractive model for intelligent databases because it provides a simple but versatile format for different kinds of knowledge, it has capabilities in artificial intelligence, and it can also function like established database models when that is required.

This paper describes how the SP model can emulate other models used in database applications and compares the SP model with those other models. The artificial intelligence capabilities of the SP model are reviewed and its relationship with other artificial intelligence systems is described. Also considered are ways in which current prototypes may be translated into an `industrial strength' working system.


\end{abstract}

\section{Introduction}

The SP theory is a new theory of computing
and cognition developed with the aim of integrating and simplifying a range of
concepts in computing and cognitive science, with a particular emphasis on
concepts in artificial intelligence. An overview of the theory is presented in \citet{wolff_icmaus_overview} 
and more detail may be found in earlier publications cited there.

\sloppy Amongst other things, the SP theory provides an attractive model for database
applications, especially those requiring a measure of human-like
`intelligence'. There is, of course, a wide variety of existing database systems 
that exhibit varying degrees and kinds of intelligence \citep{bertino_etal_2001} 
and it is reasonable to ask what may be gained by creating yet another system in 
that domain. In brief, the attractions of the SP model in this connection are that:

\begin{itemize}

\item It provides an extraordinarily simple yet versatile format for representing 
knowledge that facilitates the seamless integration of different kinds of knowledge.

\item It provides a framework for processing that knowledge that integrates 
and simplifies a range of artificial intelligence functions including probabilistic and exact forms of reasoning, unsupervised learning, fuzzy pattern recognition, best-match information retrieval,  
planning, problem solving and others.

\item At the same time, it can function like established database models, when that is required.

\end{itemize}

Prototypes of the SP system have been developed as software simulations running on 
an ordinary computer. These prototypes serve to demonstrate what can be done with 
the system and they provide the examples shown in this paper. But a programme of 
further development will be required to translate the prototypes into a 
system with `industrial strength'.

\subsection{Aims and Presentation}

The main aims of this paper are:

\begin{itemize}

\item To describe how the SP model can emulate other models used in database 
applications and to compare the SP model with those other models.

\item To review the artificial intelligence capabilities of the SP model and its relationship with other artificial intelligence systems.

\item To consider how current prototypes may be translated into a working system.

\end{itemize}

This paper does {\em not} aim to provide a comprehensive view of the SP theory and 
applications because this has already been provided in \citet{wolff_icmaus_overview} 
and earlier publications. The narrower focus of this paper is on the SP system as
an intelligent database system.

In the next section, the SP theory is described in outline. After that, 
Sections \ref{relational_section}, \ref{object_oriented_section} and \ref{hierarchical_network_section} are concerned with the first aim listed above, Section \ref{sp_intelligence_section} is concerned with the second aim and Section \ref{sp_industrial_strength} with the third.

\section{Outline of the SP Theory}

The SP theory is an abstract model of {\em any} system for processing information, either
natural or artificial. The system is Turing-equivalent in the sense that 
it can model the workings of a universal Turing machine but, 
unlike the universal Turing machine and equivalent models such as Lamda Calculus or 
Post's Canonical System, it has much more to say about the nature 
of `intelligence' \citep{wolff_1999_comp}. The entire theory is based 
on principles of minimum length encoding pioneered by Ray 
Solomonoff \citeyearpar{solomonoff_1964} and others (see \citet{li_vitanyi_1997}).

In broad terms, the system receives `New' information from its 
environment and transfers it to a repository of `Old' information. 
At the same time, it tries to compress the information as much as 
possible by finding patterns that match each other and merging or 
`unifying' patterns that are the same. An important part of this
process is the building of `multiple alignments' as described below. This provides the key to recognition, information retrieval, reasoning, learning, and other aspects of intelligence to be reviewed in Section \ref{sp_intelligence_section}.

\subsection{Representation of Knowledge}

In the SP system, {\em all} knowledge is stored as arrays of atomic {\em
symbols} in one or two dimensions called {\em patterns}. In work to date, 
the main focus has been on 1-D patterns but it is envisaged that, at 
some stage, the concepts will be generalised to patterns in two dimensions.

Although this may seem to be a very limited format, it is possible within the system 
to model a wide range of existing formats for knowledge, including context-free 
and context-sensitive grammars, condition-action rules, tables, networks and 
trees of various kinds, including class-inclusion hierarchies, part-whole-hierarchies 
and discrimination networks. Some examples will be seen below.

In the SP system, a {\em symbol} is simply a `mark' that can be matched with other symbols to decide in each case whether it is the `same' or `different'. There are no symbols in the system with `hidden' meaning such as `multiply' as the meaning of `$\times$' in arithmetic or `add' as the meaning of `+'. However, it is possible to define the meaning of any given symbol in the SP system in terms of other symbols and patterns that are associated with it in the system.

Within the system, constructs such as `variable', `value', `type', `class', `subclass', `object', `iteration', `true', `false', and `negation' are not provided explicitly. However, the effect of these constructs can be achieved by the use of patterns and symbols, and we shall see some examples below.

\subsection{Processing of Knowledge}

When any one New pattern is received, the system tries to find the best
possible match between the New pattern and one or more of the Old patterns. The result 
of this process is the creation of one or more {\em multiple alignments}, examples of 
which will be seen below.\footnote{The concept of multiple alignment in the SP 
framework is similar to that concept in bioinformatics but there are important 
differences described in \citet{wolff_icmaus_overview}.} Each multiple alignment is evaluated in terms of the principles of minimum length encoding as explained in \citet{wolff_icmaus_overview} 
and earlier papers cited there.

Figure \ref{alignment_example} is a simple example of the way in which a multiple alignment can achieve the effect of parsing a sentence, with SP patterns representing grammatical rules. By convention, the New pattern---which in this case is the sentence to be parsed---is always shown in column 0. All the other columns contain Old patterns, one pattern per column in any order. The Old patterns in this example represent grammatical rules. For example, the pattern `S NP \#NP V \#V NP \#NP \#S' in column 7 is equivalent to `S $\rightarrow$ NP V NP' in the convention of re-write rules, and the pattern `NP D \#D N \#N \#NP' in column 5 is equivalent to `NP $\rightarrow$ D N'. The entire multiple alignment divides the sentence into labelled parts and subparts like a conventional parsing and assigns a grammatical category to each word.

Contrary to what this example may suggest, the system has at least the expressive power of a context-sensitive grammar. More elaborate examples of natural language processing in the SP system and a much fuller discussion of this area of application may be found in \citet{wolff_2000}.  

\begin{figure}[!hbt]
\centering
\begin{BVerbatim}
0       1      2     3      4      5     6       7     8  

                                                 S        
                                   NP ---------- NP       
                            D ---- D                      
                            0                             
this ---------------------- this                          
                            #D --- #D                     
                                   N ----------------- N  
                                                       1  
boy -------------------------------------------------- boy
                                   #N ---------------- #N 
                                   #NP --------- #NP      
                                         V ----- V        
                                         0                
loves ---------------------------------- loves            
                                         #V ---- #V       
               NP ------------------------------ NP       
               D --- D                                    
                     1                                    
that --------------- that                                 
               #D -- #D                                   
        N ---- N                                          
        0                                                 
girl -- girl                                              
        #N --- #N                                         
               #NP ----------------------------- #NP      
                                                 #S       

0       1      2     3      4      5     6       7     8  
\end{BVerbatim}
\caption{An example of a multiple alignment that achieves the effect of parsing a sentence, with SP patterns in columns 1 to 8 representing grammatical rules.}
\label{alignment_example}
\end{figure}

In one operation, the creation of multiple alignments achieves a range of 
computational effects, depending on the kinds of Old pattern that are stored 
in the system. These effects include `parsing' (as in the example just shown), `recognition' of an unknown entity, `retrieval' of stored information, probabilistic `reasoning', logical `deduction', mathematical `calculation', and more.

In cases where the New pattern cannot be fully matched with the Old patterns, 
the system may `learn' by creating patterns that are derived from multiple alignments where partial 
matching has been achieved or, if there are no such multiple alignments, from the New pattern itself \citep{wolff_unsupervised_learning}. These system-generated patterns are added to the repository of Old patterns. Periodically, these patterns are evaluated in terms the principles of minimum length encoding and the repository of Old patterns may be purged of patterns that are least useful in those terms.

\subsection{Computer Models}

In the development of the SP theory, computer models have been created as a way of reducing 
vagueness and inconsistencies in the theory, as a way of verifying that the system really 
does work according to expectations, and as a means of demonstrating what the system can do. 
Two main models have been developed to date:

\begin{itemize}

\item SP61 which is a partial model of the system that builds multiple alignments from 
New and Old patterns \citep{wolff_2000}. This model does not attempt any learning and it does not add any 
patterns to its repository of Old patterns. All the Old patterns in the model must be 
supplied by the user when the program starts. This model is relatively stable and provides 
all the examples in this article.

\item \sloppy SP70 which is an augmented version of SP61 that builds multiple alignments and 
can learn by adding system-generated patterns to its repository of Old patterns \citep{wolff_cavtat_2003,wolff_unsupervised_learning}. This model has already demonstrated significant capabilities for learning but further work is needed to realise the full potential of the model.

\end{itemize}

\subsection{Arithmetic and Procedural Code}\label{arithmetic_procedural_section}

Since the SP system can model the operation of a universal Turing machine, it can, in principle, be used for any kind of arithmetic or mathematical operation and it can, in principle, perform any kind of `procedure' that one might program in a procedural programming language such as C++ or Cobol. That said, most applications that have been developed to date have a `declarative' flavour and the ways in which the system may be applied to arithmetic or other mathematical operations have not yet been explored in any depth (but see \citet{wolff_maths_logic}). This has a bearing on how the system may be developed for database applications, as will be discussed in Section \ref{sp_industrial_strength}.

\subsection{Computational Complexity}\label{computational_complexity}

Many problems in artificial intelligence are known to be intractable if one wishes to obtain the best possible answer. But if one is content with answers that are merely `good enough', then it is often possible to achieve dramatic reductions in time complexity or space complexity or both.

These remarks apply to the multiple alignment problem in bioinformatics and to the version of that problem that has been developed in the SP system. For any realistic example, an exhaustive search of the abstract space of possible multiple alignments is not possible and {\em constraints} must be applied, pruning away large parts of the search space. In current models, the main emphasis is on hill climbing and related techniques that concentrate search in areas that are proving productive without ruling out any part of the search space {\em a priori}---and with enough flexibility to be able to escape from `local peaks'. 

In a serial processing environment, the time complexity of the SP61 model is approximately O$(\log_2 N_s \times N_sO_s)$, where $N_s$ is the number of symbols in the New pattern and $O_s$ is the total number of symbols in the patterns in Old. In a parallel processing environment, the time complexity may approach O$(\log_2 N_s \times N_s)$, depending on how well the parallel processing is applied. The space complexity in serial or parallel environments is approximately O$(O_s)$.

In a serial processing environment, the time complexity of the SP70 model is approximately O$({N_p}^2)$ where $N_p$ is the number of patterns in New and it is assumed that they are all of the same size or nearly so. In a parallel processing environment, the time complexity may approach O$(N_p)$, depending on how well the parallel processing is applied. In serial or parallel environments, the space complexity is approximately O$(N_p)$.

\section{The Relational Model}\label{relational_section}

This section and the two that follow describe how the SP model may achieve the effect of popular database models used in `mainstream' data processing applications. This section discusses the relational model, Section \ref{object_oriented_section} is concerned with the object-oriented model and Section \ref{hierarchical_network_section} considers the hierarchical and network models.

Consider a typical table from a relational database like the one shown in Figure \ref{example_table} (from the {\em DreamHome} example in \citet[p. 80]{connolly_begg_2002}). The same information can be represented using SP patterns, as shown in Figure \ref{patterns_figure_1}.

\renewcommand*{\arraystretch}{1.25}
\setlength{\tabcolsep}{1mm}
\begin{table}[!hbt]
\scriptsize
\centering
\begin{tabular}{|l|l|l|l|l|l|r|l|}\hline
\em Staff No. & \em First Name &    \em Last Name & \em Position &  \em Sex &   \em DoB &   \em Salary &    \em Branch No. \\ \hline
SL21 &      John &      White &     Manager &       M &     1-Oct-45 &  30000   &       B005 \\
SG37 &      Ann &           Beech &     Assistant &     F &     10-Nov-60 & 12000   &       B003 \\
SG14 &      David &     Ford &      Supervisor &    M &     24-Mar-58 & 18000 &     B003 \\
SA9 &           Mary &      Howe &      Assistant &     F &     19-Feb-70 & 9000 &      B007 \\
SG5 &           Susan &     Brand &     Manager &       F &     3-Jun-40 &  24000 &     B003 \\
SL41 &      Julie &     Lee &           Assistant &     F &     13-Jun-65 & 9000 &      B005 \\ \hline
\end{tabular}
\caption{A typical table representing members of staff in a company (from \citet[p. 80]{connolly_begg_2002}).}
\label{example_table}
\end{table}

\begin{figure}[!hbt]
\centering
\begin{BVerbatim}
<staff> 0
    <staff_no> SL21 </staff_no>
    <first_name> John </first_name>
    <last_name> White </last_name>
    <position> Manager </position>
    <sex> M </sex>
    <dob> 1-Oct-45 </dob>
    <salary> 30000 </salary>
    <branch_no> B005 </branch_no>
</staff>)

<staff> 1
    <staff_no> SG37 </staff_no>
    <first_name> Ann </first_name>
    <last_name> Beech </last_name>
    <position> Assistant </position>
    <sex> F </sex>
    <dob> 10-Nov-60 </dob>
    <salary> 12000 </salary>
    <branch_no> B003 </branch_no>
</staff>)
\end{BVerbatim}
\caption{Two SP patterns representing the first two rows of the table shown in
Table \ref{example_table}.}
\label{patterns_figure_1}
\end{figure}

Readers who are familiar with XML \citep{w3c_xml_spec} will see that the way in which tables
are represented with SP patterns is essentially the same as the way in which they are
represented in XML. Each pattern begins with a symbol `$<$staff$>$' that identifies it as a pattern representing a member of staff and there is a corresponding symbol, `$<$/staff$>$', at the end. Likewise, each field within each pattern is marked by start and end symbols such as
`$<$staff\_no$>$ ... $<$/staff\_no$>$' and `$<$first\_name$>$ ... $<$/first\_name$>$'.

Unlike XML, there is no restriction on the styles of symbols that may be used. For example,
`$<$staff$>$ ... $<$/staff$>$', `$<$staff\_no$>$ ... $<$/staff\_no$>$' and `$<$first\_name$>$ ... $<$/first\_name$>$' may be replaced by symbols such as `staff ... \#staff', `staff\_no ... \#staff\_no' and `first\_name ... \#first\_name', like the symbols used in Figure \ref{alignment_example}. Any other style that is convenient may also be used. In other applications, it may not be necessary to provide start and end symbols in some of the patterns, and in some cases, start and end symbols may not be needed at all. 

At first sight, the SP (and XML) representation of a table is much more long-winded and cumbersome than the representation shown in Figure \ref{patterns_figure_1}. But a table in a relational database---as it appears on a computer screen or a computer print-out---is a simplified representation of what is stored in computer memory or on a computer disk. In relational database systems, the `internal' representation of each table contains memory pointers or tags that are close analogues of symbols like `$<$staff$>$', $<$/staff$>$', `$<$staff\_no$>$' and `$<$/staff\_no$>$' that appear in the SP and XML representations. In short, the SP representation is essentially the same as the `internal' representation of a table in a relational database. The way tables are printed out or displayed on a computer screen is largely a cosmetic matter and there is no reason why tables in an SP database should not be printed or displayed in the conventional style.

\subsection{Retrieval of Information: Query by Example}

In the SP system, the most natural way to retrieve information is in the manner of
`query-by-example'. To achieve this, patterns like those shown in Figure \ref{patterns_figure_1}
are stored as Old patterns and the query is created as a New pattern in the same format as
the stored patterns but with fewer symbols. For example, if we wish to identify all
the female staff at branch number B003, our query pattern would be `$<$staff$>$ $<$sex$>$ F $<$/sex$>$ $<$branch\_no$>$ B003 $<$/branch\_no$>$ $<$/staff$>$' or even `$<$staff$>$ F B003 $<$/staff$>$'.

Given this query as the New pattern and patterns like those in Figure \ref{patterns_figure_1}
as Old patterns, SP61 creates a variety of multiple alignments but only two of them match all the symbols in the New pattern. These two multiple alignments---shown in Figure \ref{alignments_figure_1}---identify all the female staff in branch B003, as required. As previously noted, the New pattern in any multiple alignment is always shown in column 0 and the remaining columns contain Old patterns, one pattern per column.

\begin{figure}[!hbt]
\fontsize{09.00pt}{10.80pt}
\centering
\begin{BVerbatim}
0              1                   0              1            

<staff> ------ <staff>             <staff> ------ <staff>
               4                                  1
               <staff_no>                         <staff_no>
               SG5                                SG37
               </staff_no>                        </staff_no>
               <first_name>                       <first_name>
               Susan                              Ann
               </first_name>                      </first_name>
               <last_name>                        <last_name>
               Brand                              Beech
               </last_name>                       </last_name>
               <position>                         <position>
               Manager                            Assistant
               </position>                        </position>
<sex> -------- <sex>               <sex> -------- <sex>
F ------------ F                   F ------------ F
</sex> ------- </sex>              </sex> ------- </sex>
               <dob>                              <dob>
               3-Jun-40                           10-Nov-60
               </dob>                             </dob>
               <salary>                           <salary>
               24000                              12000
               </salary>                          </salary>
<branch_no> -- <branch_no>         <branch_no> -- <branch_no>
B003 --------- B003                B003 --------- B003
</branch_no> - </branch_no>        </branch_no> - </branch_no>
</staff> ----- </staff>            </staff> ----- </staff>

0              1                   0              1

(a)                                (b)
\end{BVerbatim}
\caption{The two best multiple alignments found by SP61 with the pattern `$<$staff$>$ $<$sex$>$ F $<$/sex$>$ $<$branch\_no$>$ B003 $<$/branch\_no$>$ $<$/staff$>$' in New and patterns in Old that include patterns representing Table \ref{example_table}. These two multiple alignments are the only ones that provide a match for all the symbols in the New pattern (shown in row 0 in each case).}
\label{alignments_figure_1}
\end{figure}

Of course, there is no need for the results of the user's query to be displayed in
the manner shown in Figure \ref{alignments_figure_1}. As with the representation
of tables, there is no reason in principle why information should not be displayed
or printed in whatever format is convenient.

\subsubsection{Retrieving Information from Two or More Tables}

With relational databases, it is of course quite usual for a single query to retrieve information from two or more tables.     This subsection shows how this can be done in the SP model with an example corresponding to a simple join between two tables.

In the {\em DreamHome} example \citep[][p. 80]{connolly_begg_2002}, there is one table for clients and another for viewings of properties by clients. If we wish to know which clients have viewed one or more properties and the comments they have made (example 5.24 in \citet[p. 138]{connolly_begg_2002}), we may achieve this with an SQL query like this:

\begin{center}
\begin{tabular}{l}

SELECT c.client\_no, first\_name, last\_name, property\_no, comment \\
FROM Client c, Viewing v \\
WHERE c.client\_no = v.client\_no; \\

\end{tabular}
\end{center}

In the SP model, an equivalent effect can be achieved by creating multiple alignments like the one shown in Figure \ref{alignments_figure_2}. This is one of the five best multiple alignments created by SP61 with the pattern shown in column 0 in New and patterns corresponding to the two tables in Old. Each of these five multiple alignments shows details of a viewing (in column 1) and the client who is doing the viewing (in column 2). No other multiple alignments match all the symbols in New.

\begin{figure}[!hbt]
\fontsize{09.00pt}{10.80pt}
\centering
\begin{BVerbatim}
0                1                2            

<viewing> ------ <viewing>                     
                 11                            
                 <client> ------- <client>     
                                  6            
                                  <client_no>  
                 CR76 ----------- CR76         
                                  </client_no> 
<first_name> -------------------- <first_name> 
                                  John         
</first_name> ------------------- </first_name>
<last_name> --------------------- <last_name>  
                                  Kay          
</last_name> -------------------- </last_name> 
                                  <tel_no>     
                                  0207-774-5632
                                  </tel_no>    
                                  <pref_type>  
                                  Flat         
                                  </pref_type> 
                                  <max_rent>   
                                  425          
                                  </max_rent>  
                 </client> ------ </client>    
<property_no> -- <property_no>                 
                 PG4                           
</property_no> - </property_no>                
                 <view_date>                   
                 20-Apr-01                     
                 </view_date>                  
<comment> ------ <comment>                     
                 too                           
                 remote                        
</comment> ----- </comment>                    
</viewing> ----- </viewing>                    

0                1                2            
\end{BVerbatim}
\caption{One of the five best multiple alignments created by SP61 with the pattern shown in column 0 in New and patterns representing tables for clients and viewings in Old.}
\label{alignments_figure_2}
\end{figure}

If the system is to build multiple alignments like the one shown in Figure \ref{alignments_figure_2}, it is necessary for each pattern representing a viewing to refer to the client as `$<$client$>$ ... $<$/client$>$' rather than `$<$client\_no$>$ ... $<$/client\_no$>$' (where `...' represents `CR76' or other client number). This allows the system to access details of the client such as `first\_name' and `last\_name'.\footnote{To be fully consistent, the same idea should also be applied to the way in which a property for rent is referenced from each of the `viewing' patterns. This would mean that `$<$property\_no$>$ ... $<$/property\_no$>$' in each `viewing' pattern would be changed to `$<$property$>$ ... $<$/property$>$' and it would also mean that each of the five alignments would include a column showing the property for rent.}

\subsection{Retrieval of Information: Query Languages}\label{query_languages}

The SP system does not, in itself, provide any query language like SQL for the retrieval of information. However, the system has proved to be effective in the processing of natural languages (see \citet{wolff_2000} and Section \ref{sp_intelligence_section}, below) and there is no reason in principle why the same should not be true of artificial languages like SQL. If a query language is deemed necessary, it should be possible to specify the syntax of such a language using SP patterns and to process them within the multiple alignment framework to achieve information retrieval as required. These are matters requiring further investigation.

\subsection{Comparison Between the SP Model and the Relational Model}

One of the attractions of the relational model---and perhaps the main reason for its popularity---is the simplicity of the idea of storing all database knowledge in tables. This format is very suitable for much of the knowledge to be stored in typical data processing applications but it is by no means `universal'. It is not, for example, a good medium for representing any kind of grammar or the kinds of if-then rules used in expert systems. It can be used to represent the kinds of hierarchical structure associated with object-oriented design but it has shortcomings in this connection, as we shall see in the next section.

A major difference between the relational model and the SP model is that the SP model provides a format for knowledge that is even simpler than in the relational model. Although this simplification may seem relatively slight, it has a dramatic impact on what can be represented in the system. Many kinds of knowledge that are outside the scope of the relational model can be accommodated in the SP system and, as we shall see, it overcomes the weaknesses of the relational model in representing hierarchical structures. At the same time, it can accommodate the relational model when that is required.

The second main difference between the two models is that the relational model is designed purely for the storage and retrieval of knowledge while the SP model can, in addition, support a range of different kinds of intelligence, to be reviewed in Section \ref{sp_intelligence_section}.

\section{Object-Oriented Concepts}\label{object_oriented_section}

Since the invention of the Simula language for programming and simulation in the 1960s \citep{birtwistle_1973}, there has been a growing recognition of the value of organising software and databases into hierarchies of `classes' and `subclasses', with `inheritance' of `attributes' down each hierarchy to individual `objects' at the bottom level. An associated idea is that any object may be structured into a hierarchy of `parts' and `subparts'. These `object-oriented' concepts allow software and databases to model the structure of human concepts (thus making them more comprehensible) and they help to minimise redundancies in knowledge. And this makes it easier to modify any given body of knowledge without introducing unwanted inconsistencies. In the database world, object-oriented concepts have been developed in the `entity-relationship model' and the `enhanced entity-relationship model' (see \citet{connolly_begg_2002}) and also in a variety of `object-oriented databases' (see \citet{bertino_etal_2001}). (In the remainder of this paper, the entity-relationship model and enhanced entity-relationship model will be referred to collectively as the entity-relationship model.)

In the SP system, all the object-oriented concepts mentioned in the previous paragraph may be expressed and integrated using SP patterns, as illustrated in Figure \ref{alignments_figure_3}. As previously noted, column 0 contains the New pattern and the remaining columns contain Old patterns, one pattern per column. The order of the Old patterns is entirely arbitrary, without special significance.

\begin{figure}[!hbt]
\fontsize{07.00pt}{08.40pt}
\centering
\begin{BVerbatim}
0     1                  2                 3                4               5               

      <vehicle> -------- <vehicle> ------------------------ <vehicle>                       
      <car> ----------------------------------------------- <car>                           
      <v4>                                                                                  
      <v> -------------- <v> ------------------------------ <v>                             
      <registration> --- <registration>                                                     
LMN - LMN                                                                                   
888 - 888                                                                                   
      </registration> -- </registration>                                                    
      <engine> --------- <engine> -------- <engine> ----------------------- <engine>        
      <gasoline_type> ----------------------------------------------------- <gasoline_type> 
                                                                            spark_plugs     
                                                                            carburettor     
      <e> -------------------------------- <e> ---------------------------- <e>             
                                           <fuel> ------------------------- <fuel>          
                                                                            gasoline        
                                           </fuel> ------------------------ </fuel>         
      <capacity> ------------------------- <capacity>                                       
      2000cc                                                                                
      </capacity> ------------------------ </capacity>                                      
                                           <compression> ------------------ <compression>   
                                                                            low             
                                           </compression> ----------------- </compression>  
                                           cylinder_block                                   
                                           crank_shaft                                      
                                           pistons                                          
                                           valves                                           
      </e> ------------------------------- </e> --------------------------- </e>            
      </gasoline_type> ---------------------------------------------------- </gasoline_type>
      </engine> -------- </engine> ------- </engine> ---------------------- </engine>       
                         steering_wheel                                                     
                         <seats> -------------------------- <seats>                         
                                                            4                               
                         </seats> ------------------------- </seats>                        
                         <doors> -------------------------- <doors>                         
                                                            4                               
                         </doors> ------------------------- </doors>                        
                         <load_space> --------------------- <load_space>                    
                                                            small                           
                         </load_space> -------------------- </load_space>                   
                         <wheels> ------------------------- <wheels>                        
                                                            4                               
                         </wheels> ------------------------ </wheels>                       
      </v> ------------- </v> ----------------------------- </v>                            
      </v4>                                                                                 
      </car> ---------------------------------------------- </car>                          
      </vehicle> ------- </vehicle> ----------------------- </vehicle>                      

0     1                  2                 3                4               5               
\end{BVerbatim}
\caption{A multiple alignment created by SP61 showing how object-oriented constructs may be expressed in the SP framework.}
\label{alignments_figure_3}
\end{figure}

In this figure, column 2 contains a pattern representing the class `vehicle'. At this abstract level, a `vehicle' in this example is something with a registration number, an engine, steering wheel, seats, and so on, but the details are unspecified. Some of that detail is provided by the pattern in column 4 that represents the subclass `car'. In this example, a car is a vehicle with 4 seats, 4 doors and 4 wheels and a relatively small space for carrying luggage. Yet more detail is supplied by the pattern shown in column 1 that represents a specific instance of a car with an identifier (`v4'), a registration number (`LMN 888'), and with a gasoline type of engine with a capacity of 2 litres.

So far, we know relatively little about the engine in v4. More information is supplied by the pattern in column 3 which represents the structure of the class of internal combustion engines. At this abstract level, an engine is something with fuel, a `capacity', some level of `compression', and a cylinder block, crank shaft, piston and valves.

More detail about the engine in this vehicle is provided by the pattern in column 5 which tells us that, as a gasoline-type engine, it runs on gasoline fuel, that it has a (relatively) low compression and that, in addition to the parts mentioned earlier, it has spark plugs and a carburettor. The main alternative to gasoline-type engines is, of course, the diesel type---not shown in the figure---which runs on diesel fuel, has a (relatively) high compression, and does not need spark plugs or a carburettor.

Readers may wonder why the symbols `$<$v$>$' and `$<$/v$>$' are used in the patterns shown in columns 1, 2 and 4 and why `$<$e$>$' and `$<$/e$>$' appear in columns 1, 3 and 5. These symbols are, in effect, `punctuation' symbols that are needed to ensure that multiple alignments can be formed according to the principles described in \citet{wolff_icmaus_overview} and earlier publications.

\subsection{Discussion}

The multiple alignment concept, as it has been developed in the SP framework, provides a means of expressing all the main constructs associated with object-oriented design:

\begin{itemize}

\item {\em Classes, subclasses and objects}. In Figure \ref{alignments_figure_3}, there is a hierarchy of classes from `vehicle' at the top level (column 2) through `car' at an intermediate level (column 4) to an individual object (`v4' shown in column 1) at the bottom level. The class `engine' is also shown at an abstract level (column 3) and at a more concrete level (column 5).

\item {\em Inheritance of attributes}. From the multiple alignment in Figure \ref{alignments_figure_3} we can infer that v4 has a cylinder block, crank shaft, pistons and valves, that the engine has a low compression, that the vehicle has 4 wheels, and so on. None of these `attributes' are specified in the pattern for v4 shown in column 1. They are `inherited' from patterns representing other classes, in much the same way as in other object-oriented systems.

\item {\em Cross-classification and multiple inheritance}. The multiple alignment framework supports cross-classification with multiple inheritance just as easily as it does simple class hierarchies with single inheritance. With our `vehicle' example, it would be easy enough to introduce patterns representing, say, `military vehicles' or `civilian vehicles', a classification which cuts across the division of vehicles into categories such as `car', `bus', `van', and so on. In a similar way, vehicles can be cross-classified as `gasoline\_type' or `diesel\_type' on the strength of the engines they contain, as shown in our example.

\item {\em Parts and subparts}. In our example, the class `vehicle' has parts such as `engine', `steering\_wheel', `seats', and so on, and the `engine' has parts such as `cylinder\_block', `crank\_shaft' etc. If there was only one type of engine, then all the parts and other attributes of engines could be expressed within the `vehicle' pattern, without the need for a separate pattern to represent the engine. The reason that a separate pattern is needed---with a corresponding slot in the `vehicle' pattern---is that there is more than one kind of engine. Another reason for representing the class of engines with separate patterns is that engines may be used in a variety of other things (e.g., boats, planes and generators), not just in road vehicles.

\end{itemize}

\subsubsection{Variables, Values and Types}

It should be apparent that, in the SP system, a pair of neighbouring symbols like `$<$fuel$>$' and `$<$/fuel$>$' function very much like a `variable' in a conventional system. By appropriate alignment within a multiple alignment, such a variable may receive a `value' such as `gasoline' or `diesel' in this example. The range of possible values that a given variable may take---the `type' of the variable---is defined implicitly within any given set of patterns in Old. 

\subsubsection{Variability of Concepts}

Column 4 in Figure \ref{alignments_figure_3} shows a car as something with 4 seats, 4 doors and 4 wheels but of course we know that all of these values can vary. Sports cars often have 2 seats and 2 doors, some budget cars have 3 wheels, and a stretch limo may have many more seats and doors. In a more fully-developed example, numbers of seats, doors and wheels would be unspecified at the level of `car' and would be defined in subclasses like those that have been mentioned.

\subsection{Comparison Between the SP Model and Other Object-Oriented Systems}

Perhaps the most striking difference between the SP system and other object-oriented systems is the extraordinary simplicity of the format for knowledge in the SP system, compared with the variety of constructs used in other system---such as `classes', `objects', `methods', `messages', `isa' links, `part-of' links, and more. This subsections considers a selection of other differences that are somewhat more subtle but are, nevertheless, important. 

\subsubsection{Parts, Attributes and Inheritance}

In Simula and most object-oriented systems that have come after, there is a distinction between `attributes' of objects and `parts' of objects. The former are defined at compile time while the aggregation of parts to form wholes is a run-time process. This means that the inheritance mechanism applies to attributes but not to parts.

In the SP system, this distinction disappears. Parts of objects can be defined at any level in a class hierarchy and inherited by all the lower level. There is seamless integration of class hierarchies with part-whole hierarchies.

\subsubsection{Objects, Classes and Metaclasses}

By contrast with most object-oriented systems, the SP system makes no formal distinction between `class' and `object'. This accords with the observation that what we perceive to be an individual object, such as `our car', can itself be seen to represent a variety of possibilities: `our car taking us on holiday', `our car carrying the shopping', and so on. A pattern like the one shown in column 1 of Figure \ref{alignments_figure_3} could easily function as a class with vacant slots to be filled at a more specific level by details of the passengers or load being carried, the r{\^o}le the vehicle is playing, the colour it has been painted, and so on. This flexibility is lost in systems that do make a formal distinction between classes and objects.

Another consequence of making a formal distinction between objects and classes is that it points to the need for the concept of a `metaclass':

\begin{quotation}

\noindent ``If each object is an instance of a class, and a class is an object, the [object-oriented] model should provide the notion of {\em metaclass}. A metaclass is the class of a class.'' \citep[][p. 43]{bertino_etal_2001}.

\end{quotation}

\noindent It is true that this construct is not provided in most object-oriented database systems but it has been introduced in some artificial intelligence systems so that classes can be derived from metaclasses in the same way that objects are derived from classes. Of course, this logic points to the need for `metametaclasses', `metametametaclasses', and so on without limit.

Because the SP system makes no distinction between `object' and `class', there is no need for the concept of `metaclass' or anything beyond it. All these constructs are represented by patterns.

\subsubsection{The Entity-Relationship Model with a Relational Database}

With minor variations, the entity-relationship model has become the mainstay of data processing applications for business and administration. Diagrammatic representations of entities and relationships are normally implemented with a relational database and there are efficient software tools to do the translation, hiding many of the details. Since this combination of entity-relationship model and relational database has come to be so widely used, it will be the focus of our discussion here.

A table can be used to represent a class, with the columns (fields) representing the attributes of the class and the rows representing individual instances of the class. Each class or subclass in a class hierarchy can also be represented by a table but in this case it is necessary to provide additional fields so that the necessary connections can be made. For example, the class of `staff' in a company may be represented by a table like the one shown in Figure \ref{example_table} and separate tables may be created for each of the subclasses `manager', `supervisor' and `assistant', each of these with columns relevant to the particular subclass but not for other subclasses. In addition, each of the tables for the subclasses needs a column such as `Staff Number' so that the record of an individual in any one subclass can be connected to the corresponding record in the superclass. Similar principles apply to the division of concepts into parts and subparts.

This system works quite well for many applications but it has a number of shortcomings compared with the SP system:

\begin{itemize}

\item Using tables to represent classes means that the description of a class must always take the form of a set of variables corresponding to the fields in the table. In the SP system, it is possible to describe a class using any combination of variables and literals, according to need. It is, for example, possible to record that a vehicle has a steering wheel (as in column 2 of Figure \ref{alignments_figure_3}) without any implication that there may be alternative kinds of steering wheel to be recorded in a field with that name. It is also possible to provide a verbal description of any class, something that is outside the scope of the relational model.

\item Using tables to represent classes means that the record for every individual must have start and end tags for every field in the table regardless of whether or not that field is used. In the SP system, start and end tags are only needed for the fields that contain a value in the record for any individual.

\item The SP system allows the description of class hierarchies and part-whole hierarchies to be separated from the description of individual members of those hierarchies. By contrast, the use of tables to represent class hierarchies and part-whole hierarchies means that the structure of these hierarchies must be reproduced, again and again, in every instance. Using tables to represent either kind of hierarchy means that information that is specific to any one individual is fragmented and must be pieced together using keys. In the SP system, by contrast, information that is specific to any one individual can always be represented with a single pattern. The SP system provides for the smooth integration of class hierarchies and part-whole hierarchies in a way that cannot be achieved using tables.

\end{itemize}

\section{The Hierarchical and Network Models}\label{hierarchical_network_section}

Although the hierarchical and network models for databases have fallen out of favour in ordinary data processing applications, the network model has seen a revival, first with the development of the hypertext concept and then more dramatically with the application of that concept to the world wide web. The hierarchical model is the mainstay of hierarchical file systems and finds niche applications in directories of various kinds.

In the SP system, any network or hierarchy can be represented using connections between patterns like the connection between `engine' and `vehicle' in Figure \ref{alignments_figure_3} (columns 3 and 2). The basic idea is that one pattern, `A', may contain the start and end symbols of another pattern, `B', so that the two patterns can be connected in a multiple alignment like this:

\begin{center}
\begin{BVerbatim}
0    1      2   

     <A>        
a1 - a1         
     <B> -- <B> 
b1 -------- b1  
     </B> - </B>
a2 - a2         
     </A>       

0    1      2.
\end{BVerbatim}
\end{center}

\noindent In effect, the pair of symbols `$<$B$>$ $<$/B$>$' in the `A' pattern (column 1) are a `reference' to the `B' pattern (column 2). With this simple device, it is possible to link patterns in hierarchies and networks of any complexity. Any one pattern may appear recursively, two or more times within a multiple alignment, as described in \citet{wolff_icmaus_overview} and earlier publications.

Where the full versatility of this scheme is not needed, it is also possible to create networks and hierarchies from patterns like `$<$A$>$ ... $<$B$>$', `$<$B$>$ ... $<$C$>$' and `$<$C$>$ ... $<$D$>$' that can be linked end-to-end by alignment within a multiple alignment.

\section{The SP Model and Aspects of Intelligence}\label{sp_intelligence_section}

This section briefly reviews aspects of intelligence that have been shown to fall within the scope of the SP system, highlighting those with particular relevance to intelligent databases. The main points of difference between the SP system and other artificial intelligence systems are also reviewed. Readers are referred to \citet{wolff_icmaus_overview} and other cited sources for more detail about artificial intelligence capabilities of the system outlined here:

\begin{itemize}

\item {\em Representation of knowledge}. As previously mentioned, the format that has been adopted for representing knowledge within the SP system has proved to be remarkably versatile, despite its extreme simplicity. Given the system for forming multiple alignments, flat patterns can be used to represent context-free and context-sensitive grammars \citep{wolff_2000}, networks, trees (including class-inclusion hierarchies and part-whole hierarchies), tables, if-then rules and more. Some of this versatility has been demonstrated above.

In the context of knowledge-based systems, a benefit of this versatile `universal' format for knowledge is the scope that it offers for the seamless integration of different kinds of knowledge, minimising the awkward incompatibilities that arise in many computing systems.

\item {\em Fuzzy pattern recognition and best-match information retrieval}. At the heart of the SP system is a version of dynamic programming (see \citet{sankoff_kruskall_1983}) that allows the system to find `good' full and partial matches between patterns \citep{wolff_1994_scaleable}.\footnote{The technique that has been developed in the SP models has advantages compared with standard techniques for dynamic programming: it can process arbitrarily long patterns without excessive demands on memory, it can find many alternative matches, and the `depth' or thoroughness of searching can be determined by the user.} This allows the system to recognise objects and patterns in a `fuzzy' manner and to retrieve stored information without the need for an exact match between the retrieval query and any item to be retrieved.

\item {\em Ontologies and `semantic' retrieval of information}. The SP system provides a powerful framework for the representation and processing of ontologies and for the retrieval of information by meanings rather than literal matching of patterns \citep{wolff_semantic_web_1}.

\item {\em Analysis and production of natural languages}. The syntax of natural languages may be represented with SP patterns and both the parsing and the production of sentences may be achieved by the formation of multiple alignments \citep{wolff_2000}. Non-syntactic `semantic' structures may also be represented and processed in the SP system. Recent work, not yet published, has shown how syntax and semantics may be integrated within the SP framework.

\item {\em Probabilistic reasoning}. A major strength of the SP system is its support for probabilistic `deduction' in one step or via chains of reasoning, abductive reasoning, and nonmonotonic reasoning with default values \citep{wolff_1999_prob}. Relative probabilities of inferences may be calculated strictly in accordance with standard probability theory and the system provides an explanation for the phenomenon of `explaining away' \citep{pearl_1988}.

\item {\em Exact forms of reasoning}. Although this area is less well developed, there are good reasons to think that the SP system may also be applied to the `exact' kinds of reasoning found in many areas of logic and mathematics, where answers are either `true' or `false', with nothing in between \citep{wolff_maths_logic}.

\item {\em Planning and problem solving}. The SP system has been applied successfully to the problem of finding a route between two places \citep{wolff_icmaus_overview} and it can solve geometric analogy problems translated into textual form \citep{wolff_1999_prob}.

\item {\em Unsupervised learning}. In its overall abstract structure, the SP system is conceived as a system for unsupervised learning---and capabilities in this area have now been demonstrated in the SP70 computer model \citep{wolff_cavtat_2003,wolff_unsupervised_learning}. The results are good enough to show that the approach is sound but further development is needed to realise the full potential of this model.

If this potential can be realised, this should reduce or eliminate the need for human judgement in the normalisation of knowledge structures. The SP system should be able to organise its knowledge automatically in a way that minimises redundancies and reveals the natural structures in that knowledge, including class hierarchies, part-whole hierarchies and their integration. It should also be able to abstract rules and other generalisations from its stored knowledge, in the manner of datamining systems. Of course, existing datamining techniques may also be applied to an SP database.

\end{itemize}

\subsection{Relationship with Other Artificial Intelligence Systems}

It would take us too far afield to attempt a detailed comparison with artificial intelligence systems in the kinds of areas mentioned above. As an attempt to integrate ideas across a wide area, the SP system naturally has points of similarity with many existing systems, but at the same time, it has its own distinctive features.

Chief amongst these is the remarkable simplicity of the system combined with its very wide scope, much wider than the great majority of artificial intelligence systems, with the possible exception of unified theories of cognition such as Soar \citep{laird_newell_rosenbloom_1987,rosenbloom_etal_1993} and ACT-R \citep{anderson_lebiere_1998}. Like those two systems, the development of the SP system was inspired by the writings of Allen Newell, putting the case for greater breadth and depth in theories of cognition \citep{newell_1973,newell_1990}.

Unlike hybrid systems, of which there are many, the SP system is not merely a conjunction of two or more different systems, combining their capabilities and also their complexities. The SP system is the result of a radical rethink of concepts in artificial intelligence and beyond, aiming for integration in a radically simplified structure. The result is a conceptual framework with distinctive features of which the main ones are:

\begin{itemize}

\item {\em All} kinds of knowledge are represented with flat patterns.

\item {\em All} kinds of processing is achieved by compression of information by the matching and unification of patterns.

\item The use of a modified version of the concept of multiple alignment as a vehicle for recognition of patterns, information retrieval, probabilistic and exact forms of reasoning, and other artificial intelligence functions.

\end{itemize}

\section{Developing the System}\label{sp_industrial_strength}

The SP computer models (SP61 and SP70) are good enough to demonstrate what can be done with the system but fall short of what would be needed for applications in industry, commerce or administration. This section considers how the SP concepts that have been developed to date may be translated into a practical system.

\subsection{Parallel Processing}

Although the computational complexity of both models in a serial processing environment is within the bounds of what is normally considered to be acceptable, significant improvements are to be expected if the system can be developed with the benefit of parallel processing (Section \ref{computational_complexity}). And of course, parallel processing brings the additional benefit of faster processing in absolute terms and, with suitable design, greater robustness in the face of system failures. Parallel processing is now a recognised requirement to meet the high computational demands of large scale databases \citep{abdelguerfi_lavington_1995,abdelguerfi_wong_1998} and large-scale applications in artificial intelligence.

At the heart of the SP system is the building of multiple alignments and the core operation here is a process for finding good full and partial matches between patterns in the manner of dynamic programming. At this level, there is considerable scope for the application of parallel processing because there are often many pairs of patterns that need to be matched and this can be done in parallel just as well as it can be done in sequence. There is also scope for parallel processing at a more fine-grained level because the process of matching involves a process of `broadcasting' symbols to make yes/no matches with other symbols and this is an intrinsically parallel operation.

The SP machine does not necessarily have to be developed in silicon. One futuristic possibility is to exploit the potential of organic molecules such as DNA or proteins---in solution---to achieve the effect of high-parallel pattern matching. This kind of `molecular computation' is already the subject of much research (see, for example, \citet{adleman_1994,adleman_1998}) and techniques of that kind could, conceivably, form the basis of a high-parallel SP machine.

Another possibility is to use light for the kind of high-speed, high-parallel pattern matching that is needed at the heart of the SP machine. Apart from its speed, an attraction of light in this connection is that light rays can cross each other without interfering with each other, eliminating the need to segregate one stream of signals from another (see, for example, \citet{cho_colomb_1998,louri_hatch_1994}).

On relatively short time scales, a silicon version of the SP machine would probably be the easiest option and it may be developed in at least four different ways:

\begin{itemize}

\item It should be feasible to design new hardware for the kind of high-parallel pattern matching that is needed.

\item Given that SIMD and MIMD high-parallel computers are already in existence, an alternative strategy is to create the SP machine as a software virtual machine running on one of these platforms.

\item An existing high-parallel database system (see, for example, \citet{page_1992,mahapatra_mishra_2000}) may be modified to support the SP model. Other models may, of course be retained alongside the SP model.

\item The system may be developed using low-cost computers connected together in a LAN or even a WAN. Systems like Google have been developed in this way and they already provide high-speed pattern matching of a kind that may be adapted for use within a software implementation of the SP machine. 

\end{itemize}

\subsection{User Interface}

A graphical user interface to the SP system is needed for the input of data and queries, for the setting of parameters, and for viewing data and results. Facilities that would be useful include:

\begin{itemize}

\item The possibility of translating SP patterns like those shown in Figure \ref{patterns_figure_1} into a conventional tabular format (without start and end tags) for viewing or printing.

\item The representation of class hierarchies, part-whole hierarchies or other kinds of hierarchy or network in graphical form, without showing the tags that are used to link patterns together.

\item The ability to represent multiple alignments as flat patterns, reducing each column to a single symbol.

\item Facilities for scrolling and zooming to view any large structure such as a large multiple alignment, or a large hierarchy, network, or table.

\item Menus and dialogue boxes for controlling the system and setting parameters.

\end{itemize}

For some applications there may be a need to provide an SQL-like query language. As indicated in Section \ref{query_languages}, it seems likely that this may be achieved within the SP framework by means of an appropriate set of SP patterns---but the details of how that should be done would need investigation.

\subsection{Hybrid Solutions}

In the development and application of information systems, it is rarely possible to introduce a new model and simultaneously discard all pre-established models. There is normally a transition phase, which may be very prolonged, where two or more models coexist as alternatives for different applications or are used in some combination, according to need. As the SP system matures, it may form hybrids with other systems in at least three different ways:

\begin{itemize}

\item As noted previously (Section \ref{arithmetic_procedural_section}), the SP system is not yet a rival for well-established procedural languages like C++ or Cobol, and its application to arithmetic and other areas of mathematics needs development. However, there is no reason why the system should not be used in conjunction with existing procedural languages, and with arithmetic or mathematical functions, in very much the same way that the relational database model is standardly used in conjunction with these non-relational languages and functions in many data processing applications.

\item Although in principle the SP system is a model for any kind of software, it is likely to be some time before it would be feasible or sensible to translate all existing applications into the form of SP patterns. Meanwhile, there is no reason why an SP database system should not serve as a framework within which existing applications may be embedded, in much the same way that a relational or object-oriented database---or, indeed, an hierarchical file system---may contain pointers to executable files of many different kinds (see, for example, \citet{carino_sterling_1998}).

\item As noted previously, there is no reason why existing datamining techniques should not be used with an SP database although, in the long run, this kind of processing should fall within the scope of the SP model.

\end{itemize}

\section{Conclusion}

The SP model is an alternative to existing database models that offers significant benefits compared with those models. Within the multiple alignment framework it is possible to represent knowledge in a format that is both simple and versatile, and processing within the framework provides a key to intelligence in the recognition of patterns, retrieval of information, probabilistic and exact kinds of reasoning, planning, problem solving and others.

The versatility of the SP framework means that existing database models can be accommodated within the system and it can function in accordance with any one of those models where that is required. At the same time, it offers a range of options that are not available in systems that are dedicated to any of the existing models.

Although more work is required in understanding how the model may be developed for learning, other aspects are sufficiently robust and mature for development into an industrial strength working system. 

\section*{Acknowledgements}

I am grateful to Thomas Connolly for constructive comments on this article. The responsibility for all errors and oversights is, of course, my own.

\raggedright

\end{document}